%% file: paper.tex
\documentclass[times, 10pt,twocolumn]{article} 
\usepackage{latex8}
\usepackage{graphicx}
\usepackage{subfigure}
\usepackage{algorithmic}
\usepackage{algorithm}
\usepackage{amsthm}

\newtheorem{thm}{Theorem}

\begin{document}

\title{GeoP2P: An Adaptive Peer-to-Peer Overlay for Efficient Search and
Update of Spatial Information}

\author{Shah Asaduzzaman and Gregor v. Bochmann \\
School of Information Technology Engineering\\
University of Ottawa\\
Ottawa, ON, Canada K1N 6N5\\
\texttt{\{asad,bochmann\}@site.uottawa.ca}
}

\maketitle

\input{abstract}

\input{intro}

\input{assumption}

\input{structure}

\input{routing}

\input{maintenance}

\input{discussion}


\bibliographystyle{latex8}
\bibliography{asad}

\end{document}

%% file: abstract.tex
\begin{abstract}
This paper proposes a fully decentralized peer-to-peer overlay
structure GeoP2P, to facilitate geographic location based search and
retrieval of information. Certain limitations of centralized
geographic indexes favor peer-to-peer organization of the information,
which, in addition to avoiding performance bottleneck, allows autonomy
over local information. Peer-to-peer systems for geographic or
multidimensional range queries built on existing DHTs suffer from the
inaccuracy in linearization of the multidimensional space. Other
overlay structures that are based on hierarchical partitioning of
the search space are not scalable because they use special super-peers to
represent the nodes in the hierarchy.  GeoP2P partitions the search
space hierarchically, maintains the overlay structure and performs
the routing without the need of any super-peers. Although similar
fully-decentralized overlays have been previously proposed, they lack
the ability to dynamically grow and retract the partition hierarchy
when the number of peers change.  GeoP2P provides such adaptive features
with minimum perturbation of the system state. Such adaptation makes
both the routing delay and the state size of each peer logarithmic to
the total number of peers, irrespective of the size of the
multidimensional space. Our analysis also reveals that the overlay
structure and the routing algorithm are generic and independent of
several aspects of the partitioning hierarchy, such as the geometric shape
of the zones or the dimensionality of the search space.
\end{abstract}

%% file: intro.tex
\section{Introduction}
\label{sec:intro}
Location-based information search has become a very popular and useful
service in recent years. Almost all major web search
engines~\cite{GoogleMaps, YahooLocal} provide location-based search
tools for finding businesses and services at or around any particular
geographic location. Current implementations of these industrial
search engines rely on centralized indexing schemes maintained with
large deployment of computing and storage resources. Such centralized
architecture creates performance bottleneck for frequently updated
information, such as the number of patients currently waiting in a
clinic, or real time information such as video streams from monitoring
cameras, due to the huge aggregate volume of updates. Arrival or
departure of local entities are not readily reflected in the
centralized index either. More importantly, crawling, replicating and
indexing of all the information by a single authority denies the
autonomy of different entities over their local inforamtion. All these
requirements demand a fully decentralized self-organizing architecture,
also known as a peer-to-peer architecture, where each local entity
(called {\em peer}) maintains its own local information and indexing
is performed collaboratively without any centralized authority. Such
architecture allows autonomy over local information, removes
performance bottlenecks, distributes and balances operational load and
avoids any single point of failure.

In essence, locality-based search is realized by range queries and
nearest neighbor queries in the 2-dimensional metric space of the
earth surface. Techniques for resolving range and nearest neighbor
queries in multidimensional metric space have been extensively studied
and are well understood for centralized
databases~\cite{SpatialDB1994}. With recent advances in peer-to-peer
systems, several approaches have been proposed in the literature to
accommodate such functionalities in distributed databases with
peer-to-peer organization.

In the peer-to-peer literature, originated from the study of
decentralized file-sharing systems, peers or Internet-connected
end-hosts communicate through a self-organized network of acquaintances,
called overlay network. In a class of overlay networks, called
structured overlays, the overlay neighborhood follows a certain
pattern to facilitate efficient routing of query and update messages
to the responsible peers. Peers are usually assigned randomized unique
numerical identifiers and the structure is defined over the identifier
space which is one-dimensional in nature. They are also called
distributed hash tables (DHT) for the hash-table-like put and get
interface they provide. Like regular hash tables, DHTs are designed
for storage and retrieval of a single data item at a time and it is
hard, though not impossible, to accommodate complex queries such as
range queries in such systems. 

Nevertheless, there exist several systems that are built on top of
DHTs to accommodate range queries, even for multi-dimensional data
sets~\cite{DSTShenker2006, Dongsheng2006, FAN2008, Kantere2008}. Due
to the one-dimensional nature of the identifier space upon which the
overlay structure is built, it is relatively easier to resolve range
queries over a one-dimensional object space, if objects are hashed
into numerical keys in the same space as identifiers using a
proximity-preserving hash function. A common approach for serving
multi-dimensional range queries over the DHTs is to encode the
coordinates in multi-dimensional space into one-dimensional keys,
using space filling curves. However, all known space filling curves
have the problem of translating points of close proximity in
multi-dimensional space into relatively distant points in the single
dimension. This makes accurate resolution of the range queries harder
and inefficient.

It may be noted that DHTs work well for resolving range queries, if
the overlay structure is based on the object space, instead of the
randomly assigned identifier space. It is not usually practical to
create a customized DHT for every possible multi-dimensional object
space. However, the unique combination of widespread interest in
location-based search and the geographic distribution of information
providing peers, suggests the development of customized overlay
structures based on the geographic neighborhood.

Indeed there have been several proposals for creating overlay
structures based on geographic proximity of peers or proximity of
geographic locations represented by peers~\cite{DPTree2006,
EZSearch2008, Globase2007, DiST2006, Tanin2007, Liu2005}. The common
feature of all such proposals is that the 2-dimensional geographic
space is hierarchically partitioned into zones and the overlay
structure allows routing of the queries along the depth of the
hierarchy. Such idea of hierarchical partitioning originated from the
well-known indexing data structures for multi-dimensional data-sets
such as R-tree~\cite{RTree1984}, widely used in the realm of
centralized databases. One common problem in most of the distributed
implementations of hierarchical partitioning schemes is that they
assign special roles to some peers to represent different levels of
zones in the hierarchy. This results in peers representing higher
level zones becoming bottlenecks for query routing and single points
of failure.

In this paper, we propose a fully decentralized peer-to-peer overlay
structure named GeoP2P with hierarchical partitioning of 2-dimensional
geographic space, where the maintenance of the overlay structure and
the routing of queries are performed without any special peers in the
zones. In fact, a very similar fully decentralized overlay structure,
named P2PR-tree~\cite{P2PRTree2004} was described previously.  One
problem of P2PR-tree is that it does not properly accommodate dynamic
formation and adjustment of partitions in the presence of large-scale
peer joins and departures, or {\em churn}. The main contribution of
this paper is to show how information about dynamically formed zone
boundaries can be maintained without significant overhead, allowing
growth and retraction of the zone-hierarchy following the growth and
reduction in the number of peers. This allows us to keep both the
query routing time and the size of the state information maintained at
each peer, within an average logarithmic bound of the number of peers in the
system, irrespective of the size of the 2-dimensional space.  Our
analysis also reveals that several aspects of the zoning hierarchy
such as geometric shape of the zones or dimensionality of the space,
have minimal or no impact on the overlay structure and routing
algorithms. Moreover, we have defined detailed mechanism for
maintaining the overlay structure in the presence of churn.

The rest of the paper is organized as
follows. Section~\ref{sec:assumption} defines the problem and
clarifies our assumptions. Section~\ref{sec:structure}
introduces the overlay structure of GeoP2P based on hierarchical space
partitioning and specifies the data structure (routing table)
maintained by each peer.  Section~\ref{sec:routing} explains how
messages are routed for different types of queries, using the overlay
structure introduced before. Correctness and complexity of the routing
algorithms are also analyzed in the same
section. Section~\ref{sec:maintain} describes how the overlay
structure adapts to the peer dynamics. In particular, this section
describes how a newly joining peer initializes its routing table
(Section~\ref{subsec:maintain_bootstrap}), how the zone hierarchy is
grown and retracted with change in the number of peers
(Section~\ref{subsec:maintain_growth}
and~\ref{subsec:maintain_retract}), and how the correctness of the
routing table entries is maintained in presence of churn
(Section~\ref{subsec:maintain_recovery}). Finally, we conclude with a
discussion of our contributions compared to existing works in
Section~\ref{sec:discuss}.

%% file: assumption.tex
\section{System Model and Assumptions}
\label{sec:assumption}
The system consists of large number of {\em peers}, distributed across
a 2-dimensional space with rectangular boundary. Each peer resides in
and a point in the 2 dimensional space and responsible for providing
information relevant to that point. A peer can be a data collection
sensor such as a surveillance camera or a database regarding a
particular object related to the point such as a hotel or gas station.
The data stored in each peer is updated independently. Also, any peer
can be interested in any region in the space and launch a query. The
purpose of the overlay network is to route the query to all relevant
peers.

Although the earth surface is not 2-dimensional or rectangular, it can
be projected as a rectangular region in a 2D plane, albeit with some
distortion. Each peer is assumed to know its coordinates in the 2D
plane from some off-the-shelf method such as Global Positioning System
(GPS). Note that although we restrict our discussion in 2-dimensional
space, the the proposed scheme can be easily generalized for spaces
with 3 or more dimensions. Also, application of the proposed scheme is
not limited to geographical space, it can be used for location based
searches in virtual worlds as well as for range queries over search
spaces of multiple continuous attributes.

We assume that each peer is connected to the an underlying network
such as Internet and can potentially communicate with any other peer
in the system using transport protocols in the underlay, as long as
address of the target peer in the underlay is known. We denote the
address as {\em network address}. Peers can arbitrarily join and leave
the system. Thus it is practically impossible for a peer to know the
network address and Cartesian coordinate of all peers in the system to
resolve the query locally. Hence, the geographic space is indexed and
the query is resolved in a distributed manner. Two important design
criteria of the system are (i) to avoid assigning any special role to
any peer and (ii) to allow peers to arbitrarily join and leave the
system with minimum perturbation in the system.

The focus of this paper is to design the overlay network that
facilitates different location based queries. By `resolving queries',
we mean routing an application defined message to the peers
responsible for taking action on and/or sending reply of the
message. We leave the exact semantics of the message and response to
the application and concentrate on the routing mechanism. Example
message semantics could be sending some commands to the peers at
particular location, sending some database queries to the peers in a
location to retrieve some information regarding the location or asking
for handles for accessing some resources in the peers in a location.

%% file: structure.tex
\section{Overlay Structure}
\label{sec:structure}
In this section we discuss the structure of the GeoP2P overlay network
that routes the geographic queries to relevant peers. 

\subsection{Structured Zoning}
\label{subsec:struct_zoning}
\begin{figure*}[htb]
  \centering
  \setcounter{subfigure}{0}
   \subfigure[Zoning by clustering]{
     \label{fig:zone_clustering}
     \includegraphics[scale=0.66]{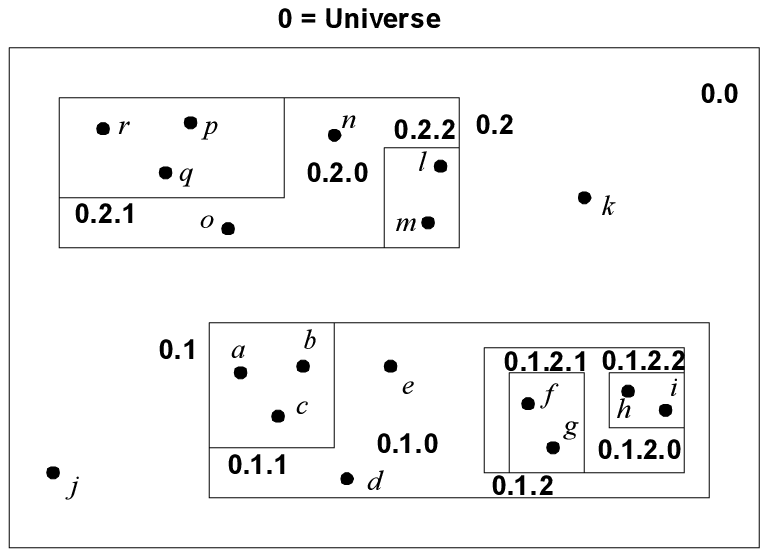}
   }
  \hspace{.1cm}
  \subfigure[Zoning hierarchy]{
    \label{fig:zone_tree}
    \includegraphics[scale=0.66]{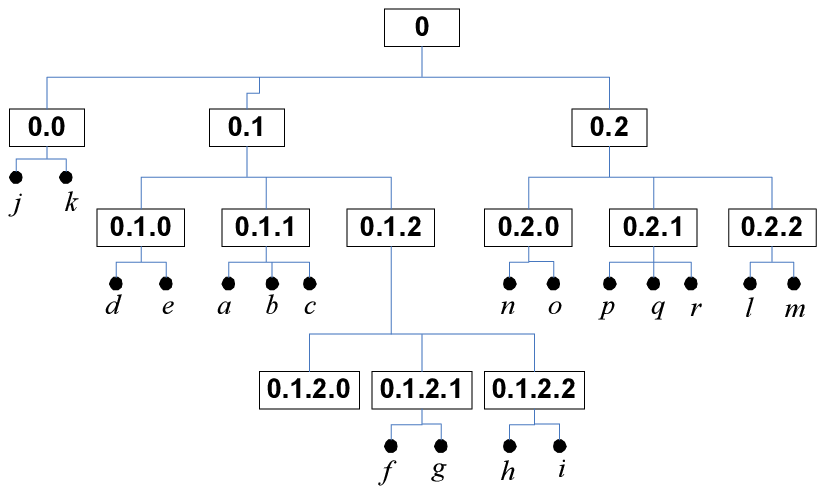}
  }
  \hspace{.1cm}
  \subfigure[Zoning by splitting]{
    \label{fig:zone_splitting}
    \includegraphics[scale=0.66]{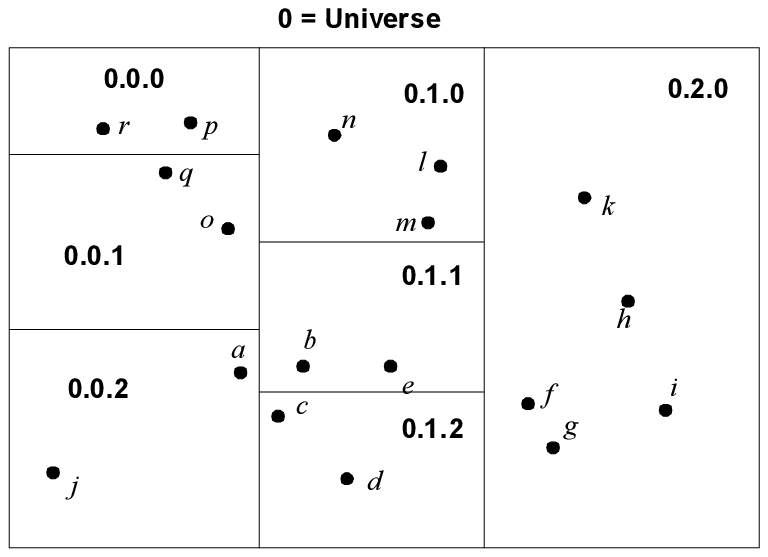}
  }
  \caption{Hierarchical zoning of the universe}
\end{figure*}

The universe is hierarchically divided into zones. At the top level of
the hierarchy, the zone representing the universe is divided into a
number of sub-zones, each of the sub-zones being further divided into
sub-sub-zones at the next level of the hierarchy, and so on. Thus the
zones can be conceptually organized into a tree, where the root of the
tree represents the universe and each tree-node represents a zone. The
zone represented by a tree-node completely contains all the sub-zones
represented by the children of that node and the zones represented by
the children completely cover the zone represented by the parent
tree-node. Also, a zone is always divided into non-overlapping
sub-zones. The leaf nodes of the tree represent the zones that are not
divided any further, which we denote as {\em leaf zone}s. Each peer
belongs to a leaf zone at its deepest level, to successively larger
zones at higher levels, and to the zone covering the universe at the
top level. Figure~\ref{fig:zone_clustering} illustrates an example
division of the universe and the corresponding tree representation is
shown in Figure~\ref{fig:zone_tree}.

The division is performed dynamically according to the number and
geographic distribution of the peers. The number of divisions and
number of hierarchical levels of divisions may grow or shrink as peers
join or depart from the system. The division of a zone into sub-zones
is motivated by the number of peers in the zone and not by the area
covered. Thus, after division, each of the sub-zones contains roughly
an equal number of peers, but the amount of area covered by different
sub-zones may vary widely. The maximum out-degree of the zoning
hierarchy, i.e. maximum number of sub-zones under a zone is limited by
a system-defined constant $k$. The number of peers in each leaf zone
is maintained roughly uniformly across the universe and kept within two
system-defined thresholds -- a higher threshold $\theta_H$ and a lower
threshold $\theta_L$. Division of a zone into sub-zones is triggered
when the number of peers in the zone is above $\theta_H$. A leaf zone
may merge with one of its sibling zones if the number of peers in the
zone goes below $\theta_L$. Here, two different
thresholds are used for triggering split and merge actions, following
the standard practice of introducing a hysteresis loop in on-off
feedback control systems. This helps avoiding oscillations between
split and merge during peer join and departure, or {\em churn}. In
fact, depending on the zoning scheme, a certain ratio of $\theta_H$ to
$\theta_L$ need to be maintained, as described later in this
section. Further details on the thresholds of split and merge can be
found in~\cite{eQuus2006}.

Division of a zone into sub-zones can be performed in two different
ways -- i) splitting or ii) clustering. The choice of the zoning scheme
is a design parameter, and only one of the zoning scheme is
followed while constructing an overlay. In the splitting scheme, a
rectangular zone is divided into $k$ rectangular sub-zones as shown in
Figure~\ref{fig:zone_splitting}. Division is always performed along
the longer dimension of the rectangle, breaking the tie in favor of
$X$-axis. In the clustering scheme, non-overlapping rectangular areas
within the zone are determined using a clustering algorithm, such
that the number of peers in each rectangle is between $\theta_L$ and
$\theta_H$ (Figure~\ref{fig:zone_clustering}). Each of the rectangular
clusters is considered a sub-zone. The remaining non-rectangular area,
which also contains some peers that are scattered and do not belong to
any of the clusters, is considered as a sub-zone too. We denote the
non-rectangular sub-zone as {\em remainder zone}. Note that a
remainder zone always remains as a leaf zone.

In both of the zoning schemes, it is possible to divide a zone into
$k$ sub-zones at a time or to offshoot one sibling at a time until the
total number of siblings at that level reaches $k$. For $k$-at-a-time
division, which we denote as {\em complete division}, $\theta_H$ needs
to be at least $k$ times of $\theta_L$, allowing a wide variation in
the number of peers per leaf zone. The other scheme, denoted as {\em
  incremental divisioning}, does not require this and is more flexible
in the sense that a zone can be divided into any number of sub-zones
between $2$ and $k$, depending on the availability and spatial
distribution of peers. This is more suitable for the clustering 
zoning scheme, where spatial distribution of peers may not be suitable
for making $k$ rectangular clusters. Incremental division, however,
requires some additional messages to perform the division, as
explained in Section~\ref{subsec:maintain_growth}.

Regardless of the choice of any particular zoning scheme, zones and
peers may be identified using a structured naming system. Strictly
speaking, naming of the zones or peers is not necessary for
construction or evolution of the overlay structure, or for the purpose
of routing messages. Such additional information may however be used
for convenience and may serve as a tool for overlay maintenance, as
described in Section~\ref{subsec:maintain_recovery}. The name would
identify the path in the zoning hierarchy from the root to the
tree-node denoting a zone or peer. Such a name can be represented by a
string of integers, which can be conveniently packed in a bit-string
of sufficient length.

\subsection{Routing Table}
\label{subsec:struct_routingtable}
Each peer maintains a {\em routing table} that lists all the other
peers it knows.  To resolve a query about any region in the universe,
a peer tries to find a peer that belongs to the leaf zones
intersecting the query region. To do that, each peer needs to have
some structured knowledge to cover the globe, such that for any zone,
it either knows all the peers belonging to that zone, or at least
knows some peer that knows more about that zone. Any query can thus be
either resolved or forwarded to a peer that has better knowledge of
the queried region.

\begin{figure*}[htb]
  \centering
  \setcounter{subfigure}{0}
   \subfigure[Overlay neighbors of peer $f$]{
     \label{fig:routing_pointers}
     \includegraphics[scale=0.72]{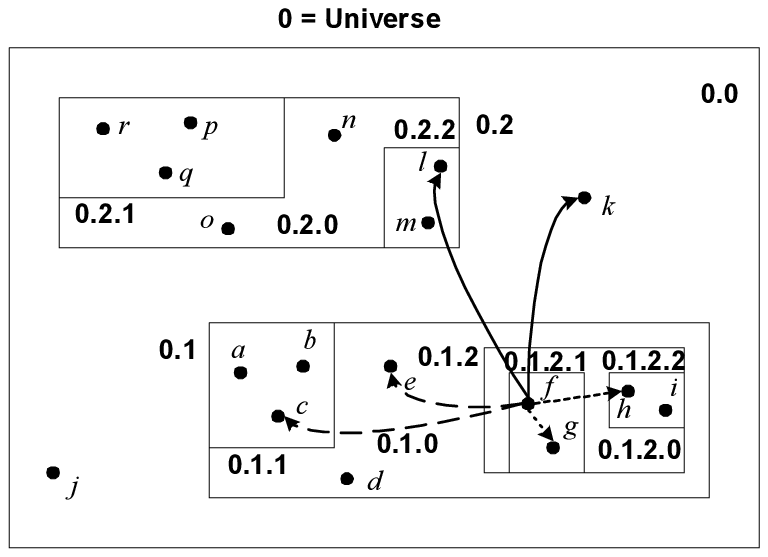}
   }
  \hspace{.1cm}
  \subfigure[Routing table of peer $f$]{
    \label{fig:routing_table}
    \includegraphics[scale=0.72]{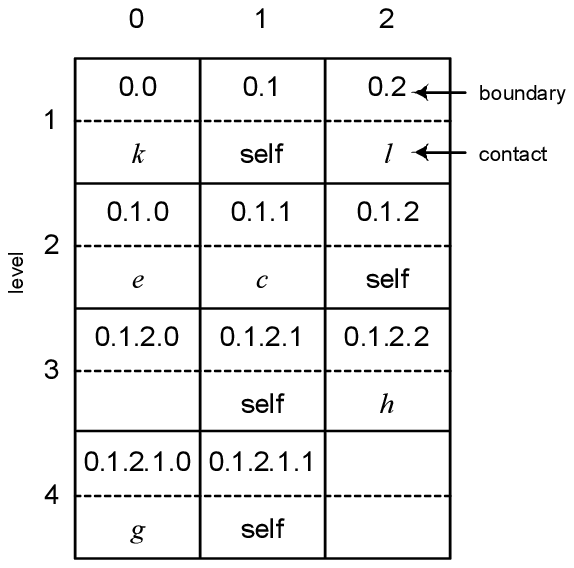}
  }
  \hspace{.1cm}
  \subfigure[Routing path of a range query]{
    \label{fig:routing_range}
    \includegraphics[scale=0.72]{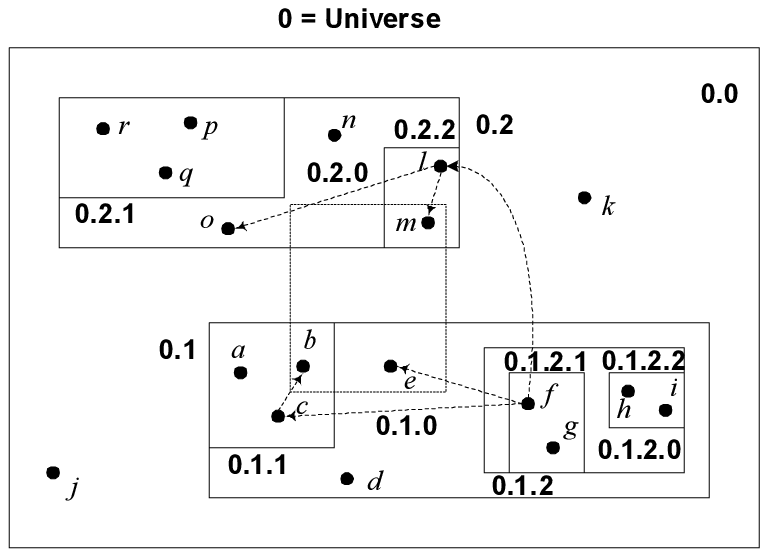}
  }
  \caption{Routing table and routing}
\end{figure*}

One way to cover the globe with minimum amount of knowledge is that,
for each level, a peer knows at least one peer in all the sibling
zones, except its own zone for that level. At the deepest level, the
peer knows all other peers within its own leaf zone. So, if a peer
belongs to a level $d$ leaf zone, its routing table contains $(k-1)d$
contact peers, where $k$ is the maximum number of divisions of a zone
at any particular level.  Assuming the tree of divisioning to be
balanced, $d=log_{k}{N}$, where $N$ is the total number of peers in
the overlay. We will demonstrate later that it is not expensive to
maintain this information correctly in presence of peer join and
departure.

The routing table is organized in $d$ rows, one for each level of
hierarchy from $1$ to $d$. Each row maintains information regarding
$k-1$ sibling zones of that level, plus some information for the
self-zone. For each sibling zone, the table need to maintain the
network address of one (or more) contact peer, rectangular boundary
(coordinates of bottom-left and top-right corner) of the
zone. Siblings can be organized into columns based on the segment of
the zone id that identifies the branching at that level. For the
self-zone, only the zone boundary need to be maintained, and it can be
stored in the corresponding column based the id of the self-zone. Level
$d$, stores the information regarding the leaf zone and here the siblings
are individual peers instead of zones. So, coordinates of the peers
are stored here instead of rectangular
boundaries. Figure~\ref{fig:routing_table} shows an example routing
table of a peer. The same overlay neighborhood is illustrated in
Figure~\ref{fig:routing_pointers}

In Section~\ref{subsec:maintain_growth}
and~\ref{subsec:maintain_retract}, we explain how the boundary
information can be retained when the network grows or shrinks due to
peer join and departure.

In a sense, the structure of the routing table in GeoP2P is very
similar to many other distributed hash tables (DHTs) such as
Pastry~\cite{Pastry2001} and Kademlia~\cite{Kademlia2002}, that use
Plaxton's prefix matching based routing~\cite{Plaxton1997}. We may
think that at each digit in the numerical identifier used in those
DHTs, from the most significant towards the least significant, the
identifier space is hierarchically divided into sub-regions, and each
node belongs to a region of the identifier space at the deepest level
of hierarchy. In that sense, the routing table of a peer in those DHTs
also maintains pointers to at least one peer in all other sibling
sub-regions at each level of the hierarchy. In fact, when zones are
identified by numbers, the same prefix-based routing works here
too. The difference between those DHTs and GeoP2P is that in our case
we have zoned the geographic space instead of the identifier space,
and also our zoning is dynamic instead of being pre-configured.

%% file: routing.tex
\section{Routing Messages}
\label{sec:routing}
As discussed in Section~\ref{sec:assumption}, the main purpose of the
GeoP2P overlay is to route messages targeted to peers in a particular
geographic location. Exact semantics of the message, which can be a
query or an update of information or a command, is left to the
application that uses the overlay. The job of the overlay is to route
the application defined messages to the specified target peers and
then {\em deliver} the message to the same application in the target.
Some of these routing services are also used by some of the overlay
management functions such as zone division or routing table
update. The target of a message may be defined in several forms, e.g.\
all peers in a specified area (used for range query), at least one
peer in a specified area or the peer closest to a specified point
(used for nearest neighbor query). The following text explains the
routing methods for different types of targets.

\subsection{Messages Targeted to Peers in an Area}
\label{subsec:func_rangeQry}
In this case a message is targeted to an area or range of interest,
and we denote them as {\em area messages}. Such messages are used for
querying information from peers in a particular area or for updating
or commanding the peers in that area. We assume that the area of
interest is specified by an axis-parallel rectangle, although it could
be any 2-dimensional geometric shape. We chose axis-parallel
rectangles, because our zoning scheme uses rectangular areas, and it
is slightly easier to decide whether a rectangular zone in the overlay
intersects with the area of interest. The message may be targeted to
all peers or at least one peer in the specified area.

\begin{algorithm}
\label{alg:rangeQuery}
\caption{Route message to all peers in area}
\begin{algorithmic}[1]
  \REQUIRE{a routing message $areaMsgAll(area, level, appMsg)$
    received by a peer $peer$ in the leaf zone $lzone$ at depth
    $d$. The message parameters are $area=$ description of the area of
    interest,
    $level=$ hierarchical level at which query need to be
    resolved and $appMsg=$ content of the applicaiton layer message} 
  
  \ENSURE{message is forwarded to all known peers
    in $area$ and to contact peers for the zones that intersect
    $area$. Message is delivered to this peer if it falls in $area$}

  \IF {$peer.coordinate$ falls in $area$}
     \STATE{Deliver $appMsg$ to $peer$} \label{algRQ:selfCheck}
  \ENDIF 
  \IF {$level \leq d$}
    \FOR {Each entry $e$ in row $d$ of the routing table}\label{algRQ:selfzoneloop_start}
      \IF {$e.zone\_boundary$ falls in $area$} \label{algRQ:cond_fallsin}
	 \STATE{Send new $areaMsgAll(area, d+1, appMsg)$ to $e.contactPeer$}
      \ENDIF
    \ENDFOR\label{algRQ:selfzoneloop_end}
  \ENDIF
  \FOR {$r=d-1$ down to $level$}
  \label{algRQ:rowloop_start}
    \FOR{each entry $e$ in row $r$ of the routing table, except for the
    one denoting self zone}
      \IF{$e.zone\_boundary$ intersects $area$}\label{algRQ:cond_intersect}
        \STATE{Send new $areaMsgAll(area, r+1, appMsg)$ to $e.contactPeer$}
      \ENDIF
    \ENDFOR
 \ENDFOR \label{algRQ:rowloop_end}
\end{algorithmic}
\end{algorithm}

Algorithm~\ref{alg:rangeQuery} summarizes the protocol that a peer
follows when it receives an area message targeted to all peers in the
specified area. Each peer forwards the message to all sibling zones at
all levels of the routing table, whose area intersects with the target
area, and to all peers within the leaf level self-zone, which fall in
the target area. While forwarding the message, the parameter $level$
is determined based on the row of the routing table in which the match
is found. This is necessary to avoid reaching the same zone from
different paths. Initially, the routing engine of the source peer
receives the message with the parameter $level=1$ from the
application. Figure~\ref{fig:routing_range} illustrates the routing
path of a message targeted to all peers in a specified area (dotted
rectangle).  Theorem~\ref{thm:routing_correctness} summarizes the
properties of the routing algorithm.

\begin{thm}
  \label{thm:routing_correctness}
  Algorithm~\ref{alg:rangeQuery} delivers an area message to all peers
  in the specified area and not to any other peer within a finite
  number of hops, as long as the routing tables are correct.  In fact
  the number of hops is bounded by $log_{k}{N}$.  Also each peer in
  the area receives the query exactly once (no redundant
  transmission).
\end{thm}

\begin{proof}
  According to the construction, for each level $l$ of the zoning
  hierarchy, the routing table of a peer maintains contact of at least
  one peer for each of the level $l$ sibling zones the peer does not
  belong to, under the peer's own level $l-1$ zone. So, if the target
  area does not intersect the zone of the current peer at level $l$,
  it can always travel to the matching level $l$ zones. If it matches
  the zone of the current peer at level $l$, level $l+1$ zones within
  this level $l$ zones are searched for match. Thus the target area is
  progressively resolved towards finer grain matching zones and the
  message is not forwarded to any zone that does not intersect the
  target area. The target area is resolved downwards for at least one
  level of hierarchy at each hop, and it never travels to zones at
  equal or higher level of hierarchy that have already been
  resolved. So the resolution terminates after $d$ hops, where $d$ is
  the maximum depth of the zoning hierarchy. According to the zone
  construction, $d = O(log_kN)$. Since only one copy of the message
  flows through each unique path of the hierarchy, the message is
  delivered to relevant peers exactly once.
\end{proof}

One may note that Algorithm~\ref{alg:rangeQuery} can easily
accommodate area definitions of any 2-dimensional geometric shape
other than axis-parallel rectangles, as long as the shape meets two
criteria -- (i) the shape can be concisely represented in the message
and (ii) there is a computationally efficient local algorithm to
decide whether a rectangular zone intersects (needed in
Line~\ref{algRQ:cond_intersect}) or a point falls in the specified
region (needed in Line~\ref{algRQ:cond_fallsin}). For example, a
circular shape meets both the criteria. It can be represented with the
center and the radius parameters, and it is not computationally hard
to decide the intersection and the falls-in conditions. Moreover,
messages with a circular target area are useful, e.g.\ to find a peer
within certain distance or to find the nearest peer
(Section~\ref{subsec:func_coverQry}). In fact, following the same
argument, the zoning hierarchy itself can be defined based on zones of
any shape other than rectangles.

When the message is $areaMsgAny$, which is destined for any peer in
the area, instead of all peers, Algorithm~\ref{alg:rangeQuery} can be
applied with simple modifications. The procedure terminates after
Line~\ref{algRQ:selfCheck} if the current peer is a matching peer. The
loop in
Lines~\ref{algRQ:selfzoneloop_start}-~\ref{algRQ:selfzoneloop_end}
terminates as soon as a matching peer is found. The rest of the
algorithm remains the same.

A special variant of area message is a message with a target area
defined by a zone in the zoning hierarchy. Such zone broadcasting is
used by some of the overlay management operations described in
Section~\ref{sec:maintain}. Although the same
Algorithm~\ref{alg:rangeQuery} can be used for this purpose, for
efficiency it may be implemented by avoiding the area intersection
conditions.  Algorithm~\ref{alg:broadcast} summarizes the modified
protocol to forward a message targeted to all peers within the
level $l$ self-zone of the peer.

\begin{algorithm}
\label{alg:broadcast}
\caption{Route message to all peers in self zone}
\begin{algorithmic}[1]
  \REQUIRE{a message $zoneBroadcast(level, appMsg)$ received by a peer
    $peer$ that resides in the leaf zone $lzone$ of depth $d$.}
  \ENSURE{Message
    is forwarded to all known peers in the self leaf zone of the
    $peer$ and to all the contact peer responsible for a zone that are
    contained in the self zone at level $level$.}  
  \STATE{Deliver $appMsg$ to $peer$}
  \IF {$level \leq d$}
    \FOR {Each entry $e$ in row $d$ of the routing table}\label{algBC:selfzoneloop_start}
      \STATE{Send new $zoneBroadcast(d+1)$ to $e.contactPeer$}
    \ENDFOR\label{algBC:selfzoneloop_end}
  \ENDIF
  \FOR {$r=d-1$ down to $level$} \label{algBC:rowloop_start} 
    \FOR{each entry $e$ in row $r$ of the routing table, 
        except for the one denoting self zone}
      \STATE{Send new $zoneBroadcast(r+1)$ to $e.contactPeer$}
     \ENDFOR
  \ENDFOR\label{algBC:rowloop_end}
\end{algorithmic}
\end{algorithm}

\subsection{Message Targeted towards a Point}
\label{subsec:func_coverQry}
Messages whose target is defined by a point, denoted as {\em point
  message}, may have several types of targets which are useful for
  different purposes. One possible target is a peer closest to the
  specified point. Another target of interest would be all or any peer
  in a leaf zone where the specified point falls in. Because it
  matches with the overlay structure, routing to peers in the same
  leaf zone of the target point is easier than routing to nearest
  peer. This message is useful when a new peer wants to join the
  overlay. It can be routed in almost the same way as the area message
  is routed using Algorithm~\ref{alg:rangeQuery}. Only the loop in
  Lines~\ref{algRQ:rowloop_start}-\ref{algRQ:rowloop_end} terminates
  as soon as one matching zone is found, because a point cannot
  intersect more than one zones. In case any one peer is sought
  instead of all peers in the leaf zone, the loop in
  Lines~\ref{algRQ:selfzoneloop_start}-\ref{algRQ:selfzoneloop_end}
  terminates as soon as one matching peer is found.

Routing a message to the nearest peer is little bit more complex than
routing to any peer in the same leaf-zone of the point, because, the
nearest peer may not reside in the same leaf zone. Routing of this
message is done in two steps. First, using the same technique as
described in previous paragraph, the message can reach at least one
peer in the leaf zone that contains the specified point. Since this
peer knows coordinates of all the peers in the leaf zone, it can
determine the in-zone candidate peer that is closest to target
point. To determine if any other peer exist in the universe which is
closer to the target point, the current peer performs a range search
in the circular area centered at the taget point and radius equal to
the distance of the in-zone closest peer from the point. Such search
is performed by sending an area-taregetd query message asking the
peers to respond with their coordinates and network addresses. After
receiving the response, the current peer can determine the peer
closest from the taeget point in the universe, and forward the message
to that peer for delivery.

%% file: maintenance.tex

\section{Maintaining the Overlay Strucutre with Peer Dynamics}
\label{sec:maintain}
\subsection{Peer Join and Routing Table Creation}
\label{subsec:maintain_bootstrap}
When a new peer wishes to join the overlay it needs to initialize its
routing table to get connected. We assume that before joining the
overlay, the new peer knows its own network address and coordinate,
and the network address of some peer already participating in the
overlay. First, the new peer needs to find the leaf zone where its
coordinate belongs to. To do this, the new peer sends a join message
targeted to any peer in the leaf zone that contains its coordinate
(Section~\ref{subsec:func_coverQry}). The peer that receives the join
message informs all other peers in the same leaf zone of the existence
of the new peer. It also replies back to the new peer with a copy of
its own routing table. This table is a valid routing table for the new
peer, except for adding the peer that replied the join message, in the
leaf zone.

After copying the routing table in the way described above, the new
peer is able to both route and receive messages. However, for the
purposes of both reliability and load balancing, it is important to
have diversity in the routing tables among the peers in the same
zone. To achieve this, the new peer asks each peer in its routing
table except those in the last row (peers in the same leaf zone), to
send back copies of their routing tables. Say $T$ is the routing table
sent back by the contact peer at row $r$ and column $c$ of the initial
routing table. Any entry of $T$ found in any row greater than $r$ is a
valid entry for the row $r$, column $c$ of the routing table of the
new peer. A random sample of these entries may be used by the new
peer. In fact, to reduce message size, random sampling is performed at
the contact peer, with the row and sample size being specified when
the sample is sought.

Is the existence of the new peer known to the rest of the universe?
Right after join, all the peers in the same leaf zone stores the
address and coordinate of the new peer in their last row of the
routing table. Also, when the new peer contacts other peers for
diversification of its routing table, those peers also become aware of
the new arrival. They actually store the address of the new peer in
their routing table, because the routing table entries may be
refreshed whenever a message is received, as described later in
Section~\ref{subsec:maintain_recovery}. The new peer will eventually
be known to the remainder of the universe too, either due to the
application messages it will generate or due to the periodic refresh
performed by every peer.

\subsection{Network Growth and Zone Creation}
\label{subsec:maintain_growth}
As mentioned before, an leaf zone is divided into sub-zones, when the
number of peers in the leaf zone grows above the higher threshold
$\theta_H$. The task of dividing a leaf zone can be performed by any
peer within the zone.  According to the construction of the routing
table, each peer in the zone is aware of the zone-boundary before the
division. Each peer is also aware of the coordinates and network
addresses of all other peers inside the leaf zone, as well as the
total number of peers in the zone. Since any peers would follow the
same zoning algorithm using the same input information, they would
result in the same zone division. However, to avoid any inconsistency
due to network dynamics during the time when divisioning is done, the
operation is performed by exactly one of the peers in the zone. To
ensure this, the peer that first detects the necessity of dividing a
leaf zone, invokes a simple one-round leader election protocol among
all the peers within the zone, where the tie is broken in favor of
higher numerical value of the network address. After performing the
division, the leader communicates the new sub-zone boundaries to all
peers within the zone before division. Each peer can now decide which
of the sub-zones it belong to, based on the sub-zone boundaries and
its own coordinate.

Each peer updates its routing table based on the new
information. After division of a level $d$ zone into $k$ sub-zones of
level $d+1$, each peer needs to update its routing table entries for
level $d$ and $d+1$. The new level-$(d+1)$ entries will be a subset of
the previous level-$d$ entries, pointing to only those peers that are
located within the same level-$(d+1)$ zone. $k-1$ entries from the
remainder of the previous level-$d$ entries will fill the level $d$ of
the new routing table. Because the boundaries of the other $k-1$
level-$(d+1)$ zones are known, the peer can randomly choose one of the
previously known peers for each of these zones. In addition, zone
boundaries of those $k-1$ zones are stored in these entries, instead
of the point coordinates previously stored. The remaining entries in
the routing table can be discarded, or, for reliability purpose, can
be stored as backup entries, as discussed later in
Section~\ref{subsec:maintain_recovery}.

From the procedure discussed above, it is obvious that the routing
tables of all the peers can be updated very easily in only one round
of message exchange, transmitting only $\theta_H-1$ messages at
most. The content of the messages is also very small, only the
boundaries of the $k$ newly formed zones need to be communicated. The
routing table of only those peers that belong to the divided zone need
to be updated, the maximum number of which is $\theta_H$. Peers outside
the zone are not affected.  The computation done at each peer is also
very simple and perturbs only the last two rows of the routing
table. To minimize alteration of the routing tables, zone boundaries
are not modified once zones are created. The only permitted ways to
adjust the number of peers in a zone are dividing into sub-zones or
merging with sibling zones (as discussed below).

\subsection{Peer Departure and Network Retraction}
\label{subsec:maintain_retract}
Besides growing due to newly joining peers, the overlay may also shrink
in size due to departure of peers. It is useful to contract the zoning
tree along with this shrinkage of the network, so that the number of
routing table entries are reduced accordingly. Reducing the depth of
the zoning tree also reduces the number of overlay hops needed for
message routing. One way of contracting the zoning tree is to merge a
leaf zone that has very low number of peers with some other zone.

Merger may be triggered when the total number of peers in a leaf zone
goes below the lower threshold $\theta_L$. The question is with which
zone to merge. Because the peers in the merging zone know about the
boundary and at least one peer of each of the sibling zones, those
zones become natural candidate for being merging partner. From the
boundaries of all the siblings, one zone can be selected such that
the resulting zone after merger will be a continuous rectangular
region. This restriction may be relaxed, i.e. the merged zone need not
be rectangular, in case a clustering zoning scheme is used and the
merger is done with the remainder zone. We denote the zone that
initiates the merger as {\em merging zone} and the zone that is chosen
as merging partner as {\em partner zone}.

The merger is simple if the partner zone still is a leaf zone,
i.e. no further split has occurred in it. Any peer in the merging zone
may initiate the merger, we denote it as {\em initiator}. To avoid
concurrent merger initiations, a the initiator performs leader
election among the peers in the leaf zone, in the same way as done
during zone divisions. Hereafter, we denote the winner of the election
as {\em initiator}. The initiator knows at least one peer in the
partner zone (denoted as {\em partner peer}). The partner peer knows
all other peers in the partner zone. The initiator sends the {\em
  merger request} to the partner peer. The request contains the
boundary of the merging zone and address and coordinates of all peers
in that zone. On receiving the request, the partner peer realizes that
it needs to extend the boundary of its own leaf zone and include the
peers given in the message as neighbors. Say both the merging and the
partner zones are at level $d$ of the hierarchy. So, the partner peer
also need to update its level $d-1$ of its routing table, by removing
the entry corresponding to the merging sibling.

Besides updating its own routing table, the partner peer also
forwards the merger request to all other peers in its zone, so that
all of them make the similar update in their routing tables.  The
initiator on the other hand, need to send a {\em merger update}
message to each of the other siblings (except the merging partner) so
that peers in them remove the merging zone from level $d-1$ of their
routing tables. The initiator knows at least one peer in each sibling
zone, so it can transmit the message to the known peer, which in turn
can broadcast the message to all peers in its zone. Lastly, the
partner peer need to respond to the initiator peer, with the address
and coordinates of all other peers of the partner zone, so that all
other peers in the merging zone can add them to their level $d$
entries in the routing table.

In case the partner zone is not a leaf zone, there need to be some
additional steps of information propagation. The partner peer, on
receiving the merger request from a level $d$ sibling, will realize
that it no longer belongs to a leaf zone at level $d$. It may belong
to a leaf zone at level $d+x$. It then broadcasts a zone-collapse
request to all the peers in its own level-$d$ zone. After collapse is
complete, the contact peer performs the rest of the merger procedure as
described before.

The zone collapse request asks for collapsing all the zoning beyond
level $d$. On receiving the collapse request, each peer reply back
with a collapse accept message. The reply message includes the network
address and coordinates of all peers in the self leaf zone of the
responding peer. The peer that requested the collapse, then aggregates
the peer information and sends back that aggregated information to
every responding peer with an announcement of the completion of the
collapse.

Note that if the clustering zoning scheme is used instead of
splitting, then the remainder zone will always be there as an leaf
sibling zone at each level. Also the remainder zone does not need to
remain rectangular and it can form a continuum with any of its
siblings. So, in case of clustered zoning, merging is always done with
the remainder zone. The routing table entry for the remainder zone is
always maintained even if there is no peer to represent that zone.

\subsection{Refreshing Routing Table Entries}
\label{subsec:maintain_recovery}
As we described the routing table structure in
Section~\ref{subsec:struct_routingtable}, each peer maintains contact
of at least one peer for all the sibling zones at each level of the
zoning hierarchy. In presence of node join and departure, it is quite
possible that the peer that was chosen as a representative contact
during last update of the routing table is no longer there. So there
is a need for continuous refreshing of the routing table entries,
particularly of the representative contact peers. We avoid
modification of the zone boundaries other than those during merger or
division, so zone boundaries need not be refreshed continuously.

One way of refreshing the entries is to use the existing application
traffic. If peer A forwards any message to peer B at level $d$, B
being an entry in level $l$ of A's routing table, then peer A is also
a valid entry for level $l$ in B's routing table. To have peer A as a
routing table entry, peer B needs three information -- network address
of A and the row and column of the routing table where A would fit in.
We may assume that whenever a message is sent from a source to
destination through the underlying network, the message is tagged with
the source address, which is true for Internet protocol. Also, as we
saw in the algorithms described in Section~\ref{sec:routing}, every
routed message has a level information, and this directly corresponds
to the routing table row where the update will be done. To resolve the
column, we need that a peer tags the message with its own zone id
whenever it forwards a message. The segment of the zone-id that
identifies the zone at level $l$, corresponds to the column of the
routing table entry to be updated. This in turn implies that, a peer
does not need to tag messages with its fully-qualified zone-id, only
the segment of the id that corresponds to the message level would
suffice.

When storing some extra bits of information is not expensive, as is
the case in present day desktop computers, a bucket of peer-addresses
can be stored for each entry of the routing table for reliability
purpose. The most recently seen peer in the bucket would be used for
routing. To implement this policy, the peers in a bucket are
maintained in an ordered list with most recently seen peer at the
front. Whenever a new active peer is discovered, it is added in the
front of the list and the peer at the tail is discarded in case of
bucket overflow. Whenever an existing peer is found to be active, it
is brought to the front.  Similar techniques have been used by
Kademlia DHT~\cite{Kademlia2002}, which splits the id-space in a
binary hierarchy.

One may note that the routing table updates using application messages
is not sufficient to maintain the correctness of the routing
table. Some peers may not generate or forward any message for a long
period. Also, the knowledge of a peer-departure is not disseminated in
this way. For this reason, each peer periodically refreshes its
routing table by explicit message exchanges. The refresh mechanism
maintains the invariant that at least one peer in each bucket is seen
within the last $t$ units of time. To aid the implementation, the
timestamp of the most recently seen peer is maintained for each
bucket. This timestamp is checked for all the routing table entries at
every $t/2$ units of time. If the elapse time from the timestamp is
found to be more than $t/2$ for any entry, an explicit refresh is
initiated for that.

To refresh an entry of row $r$ and column $c$ 
an echo message (ping) is sent to the most recently seen peer. If no
response is found within a short timeout period, this is
repeated for the subsequent peers in the bucket in order of recency,
until one of them responded. The responding peer is moved to the front
of the list and the timestamp of the bucket is updated.

In case no peer in the bucket responded, the most recently seen
contacts of each of the sibling zones of row $r$ are asked to send its
row $r$ column $c$ entry. If no new peer is discovered or no other
sibling exist, each of the peer in the same leaf zone is requested for
its routing table entry of row $r$ and column $c$. If no new peer is
discovered even in this phase, the zone corresponding to the routing
table entry is considered out of contact until next refresh. In other
cases, when some new peers are discovered, they are contacted
sequentially to verify their liveness, until one of them responds. To
maintain the routing table diversity, while contacting, the peer is
also requested to send a random sample of its routing table entries of
all rows higher than $r$. The contacted peer uses only the most
recently seen peer in each of its buckets in the sample.  The
refreshing peer stores the sample in its bucket after the responding
peer. The timestamp is set to current time.

For the entries of the last row, which are peers within the same leaf
zone, each bucket contains exactly one peer. Their existence is also
verified by echo messages at every $t/2$ unit of time, excepting those
which sent some message within last $t/2$ time units. The peers that
did not respond to the echo within the short timeout period, are
considered to have departed. This allows detection of peer departure
within the same leaf zone. The knowledge of peer departure is
eventually spread to the rest of the world, due to the refresh
mechanism. Also, when one peer departs gracefully without crashing, it
may inform all its contacts in the routing table.

The message overhead of the explicit refresh mechanism is proportional
to the bucket size and inversely proportional to the refresh period,
both of which are system design parameters. Appropriate bucket size
and period may be determined empirically, based on the typical
messaging frequency, message source distribution and reliability
requirement of a particular application.

%% file: discussion.tex
\section{Discussion}
\label{sec:discuss}
In this section we explain the distinct features of the overlay
structure and indexing scheme of GeoP2P compared to other known
schemes designed and used for similar purposes.

\subsection{Adaptive Hierarchical Zoning with Super-nodes}
The purpose of our work was to design a fully decentralized
peer-to-peer system based on hierarchical space partitioning, that
would not rely on super-peers for supporting the area hierarchy, as
used in other overlays such as Globase~\cite{Globase2007} and
EZSearch~\cite{EZSearch2008}. Globase uses the clustering based approach
the zone hierarchy, which gives more flexibility of space partitioning
when geographic distribution of peers is non-uniform. However, Globase
assigns nodes in the area hierarchy to special super-peers, which
makes it less scalable. Because, the super-peers supporting the area
nodes near the top of the area hierarchy may get overloaded from
handling search queries. Also, failure of higher level super peers may
result in large scale network partitioning. For these reasons,
explicit reliability and load balancing techniques e.g.\ back-up
super-peers become necessary in such systems.

\subsection{P2PRTree: Hierarchical Zoning without Super-node}
P2PR-tree~\cite{P2PRTree2004} is another hierarchical space portioning
based peer-to-peer overlay that does not rely on any special nodes.
However, it assumes a special kind of area hierarchy: the first two
levels of the hierarchy are defined in the form of an R-tree by a
pre-defined static grid; the lower levels of the hierarchy are
dynamically grown depending on the number of peers that will enter
into the different sub-areas. For this dynamic part of the tree, a
kind of clustering approach is chosen. However, the paper remains
vague about the question how a new peer not falling into one of the
clustered sub-areas is integrated into the tree. Also the possibility
of loosing peers and the possible retraction of the area hierarchy is
not considered in that paper.

\subsection{Systems using Space Filling Curves and Common DHTs}
There is another class of indexing systems that are built over
existing DHT overlays to serve range queries in multidimensional
object space~\cite{DSTShenker2006, Dongsheng2006, FAN2008,
  Kantere2008, PHT2004, Placelab2005, Schmidt2004}. A common feature
of all these systems is that they leverage the use of existing
well-known DHT overlay structures. In DHTs, peers are named using
identifiers randomly chosen from a linear or one-dimensional numeric
space and the overlay structure is based on the numeric proximities of
these identifiers. Such structure allows range searches in
one-dimensional space. What all the abovementioned systems commonly do
is that they translate a range (or a point) in the multidimensional
object space into one dimensional numeric space using a space filling
curve (SFC).

\begin{figure}
\centering
\includegraphics[scale=1.0]{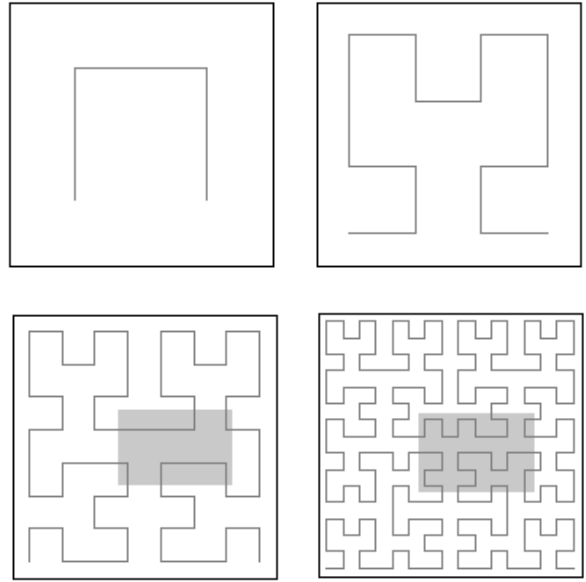}
\caption{Hierarchical property of Hilbert space filling curve in
  2-dimension. Shaded area shows a range-query target that spans
  discontiguous fragments of the curve}
     \label{fig:hilbert_hierarchy}
\end{figure}

One major problem in this translation is that a contiguous range in
multi-dimensional space does not translate into a contiguous range in
the SFC in most cases. To solve this problem, the hierarchical
property of the SFC is exploited. SFC maps the multi-dimentional
universe by mapping its different segments in gradually finer
granularities. If the universe is a $d$-dimensional hypercube, it is
divided into $n^d$ equal sized sub-cubes or cells. An approximation of
SFC is obtained by joining the centers of these cells in a poly-line
where each cell is connected to two adjacent cells. The part of the
SFC representing each cell can be expanded further into a finer grain
SFC. Thus after expansions up to depth $k$, the resulting SFC contains
$n^{kd}$ segments, covering equal number of level-$k$ cells
(Figure~\ref{fig:hilbert_hierarchy}). Now, if the cells are identified
using the distance along the SFC from its starting point in base-$n$
numbers, then the indentifiers that refer to the points in the same
level $k$ cell have a common $(k-1)d$ digit
prefix~\cite{Schmidt2004}. DHTs allow looking up any one or all peers
having certain identifier prefix. So, to route a query message to a
multi-dimentional range target in DHT, the range is hierarchicaly
translated into segments of the SFC, as the message is routed to the
peers with prefix matching to the current segment.

In the abovementioned DHT based systems, the overlay structures are
defined on identifier space. As the peer identifiers are assigned at
random, the overlay neighborhood does not have any correlation with
the physical proximity of the peers. This may be a rational choice
when the overlay is to be used for search in different
multidimensional object-spaces. We argued however that the geographic
distribution of information providing peers and the uniquely high
demand in location based search by many applications justify
construction of a special purpose overlay for the 2-dimensional
geographic search space. It would then be efficient, if the overlay
neighborhood is chosen based on the geographic proximity of the peers.
It is possible to map the geographic coordinate of the peers into
numeric identifiers using a SFC and to construct a Pasrty-like DHT
using the identifiers~\cite{ManiyPhDThesis}. The resulting overlay
structure would then be the same as GeoP2P overlay, if the hierachical
zoning is performed along a static grid like the one used for
hierarchical derivation of the SFC. In this sense, the GeoP2P overlay
is a generalization of the SFC based overlay structure, where, instead
of a fixed zoning scheme, zones of arbitrary size and shape can be
created adaptively based on the geographic density of the peers.

Note that although the systems like PHT~\cite{PHT2004} or
Squid~\cite{Schmidt2004} have been applied for range search in the
2-dimensional geographic space~\cite{Placelab2005}, their overlay
structures are based on randomly assigned identifier space and thus
have a fundamental difference with the overlay structure of GeoP2P. To
perform 2-dimensional spatial search, they map the queried 2-d range
into gradually fine grained segments of the SFC and look up the
numeric keys resembling those SFC segments in the underlying DHT. 
 
\section{Summary}
Although the peer-to-peer overlay structure and the decentralized
spatial indexing scheme presented in this paper has many similarities
with the other existing approaches, it contains several unique
features. In summary, the main contributions made by this paper are
the following:

First, we have described a generalized overlay structure based on
hierarchical space partitioning, and demonstrated that certain aspects
of the zone hierarchy have only minimal impact on the data structure
and algorithms required for maintaining the overlay and routing
different types of queries. In particular, the following aspects do
not have any major effect on the structure and function of the
overlay: (a) space partitioning scheme such as clustering or
splitting, (b) dimensionality of the universe (c) geometry of the
zones and the query region (circular or rectangular) and (d) peer
representing a point or an area in the universe

Second, we have defined detailed algorithms for query routing. In
addition to the standard range query routed to all peers associated
with the search area, we have defined algorithms for routing messages
to a single peer in the area, or to the peer that is closest to a
given point in space.

Third, we have described detailed procedure for maintaining the
virtual zoning hierarchy in the presence of churn. In addition to
explaining how the virtual zoning hierarchy may grow when the number
of peers in a given zone increases, we have also explained how the
hierarchy may retract when the number of active peers decreases.

Although we allowed dynamic growth and retraction of the zone
hierarchy in GeoP2P, we did not consider modifying the zone boundaries
once a zone is created. Allowing such modifications will provide more
flexibility in zoning when the spatial distribution of peers rapidly
changes. In future we will study how to maintain the GeoP2P overlay
structure in presence of such modifications without large scale
propagation of information. Maintaining the overlay structure in
presence of mobile peers is another related issue, which also remains
to be considered in the future.